\documentclass[12pt]{article}

\usepackage{amsmath,amssymb,epsfig,fleqn,url,psboxit,color,here,graphics,graphicx,rotating,multirow,wrapfig}

\usepackage{hyperref}
\hypersetup{
    colorlinks,
    citecolor=blue,
    filecolor=blue,
    linkcolor=blue,
    urlcolor=blue
}

\def\Title#1{\begin{center} {\Large #1 } \end{center}}
\def\Author#1{\begin{center}{ \sc #1} \end{center}}
\def\Address#1{\begin{center}{ \it #1} \end{center}}

\newcommand\pubblock{\rightline{\begin{tabular}{l} \pubnumber\\
         \pubdate  \end{tabular}}}
\newenvironment{Abstract}{\begin{quotation}  }{\end{quotation}}
\newenvironment{Presented}{\begin{quotation} \begin{center}
             PRESENTED AT\end{center}\bigskip
      \begin{center}\begin{large}}{\end{large}\end{center} \end{quotation}}

\def\beq{\begin{equation}}
\def\eeq#1{\label{#1}\end{equation}}
\def\eeqn{\end{equation}}

\def\beqa{\begin{eqnarray}}
\def\eeqa#1{\label{#1}\end{eqnarray}}
\def\eeqan{\end{eqnarray}}

\let\bar=\overbar

\def\Dslash{\not{\hbox{\kern-4pt $D$}}}
\def\dslash{\not{\hbox{\kern-2pt $\del$}}}

\def\msb{{\bar{\ssstyle M \kern -1pt S}}}

\newcommand{\slim}{\mskip 1.5mu}              

\newcommand{\phiH}{\phi _h}
\newcommand{\phiS}{\phi _S}

\newcommand\pubnumber{arXiv:1511.09093}
\newcommand\pubdate{\today}

\begin{document}
\begin{titlepage}
\pubblock

\vfill \Title{Transverse spin azimuthal asymmetries in SIDIS at COMPASS: Multidimensional analysis} \vfill \Author{
Bakur Parsamyan\\ (on behalf of the COMPASS collaboration)} \Address{University of Turin and Torino Section of INFN\\
 Via P. Giuria 1, 10125 Torino, Italy }
\vfill
\begin{Abstract}
One of the important objectives of the COMPASS experiment (CERN, SPS north area) is the exploration of transverse spin
structure of nucleon via study of spin (in)dependent azimuthal asymmetries with semi-inclusive deep inelastic
scattering (SIDIS) processes and recently also with Drell-Yan (DY) reactions. In the past twelve years series of
measurements were made in COMPASS, using 160 GeV/c longitudinally polarized muon beam and polarized $^6LiD$ and $NH_3$
targets. Drell-Yan measurements with high energy (190 GeV/c) pion beam and transversely polarized $NH_3$  target
started in 2014 with a pilot-run have been followed by 140 days of data taking in 2015. The experimental results
obtained by COMPASS for azimuthal effects in SIDIS play an important role in the general understanding of the
three-dimensional nature of the nucleon and are widely used in theoretical analyses and global data fits. In addition,
future first ever polarized DY-data from COMPASS compared with SIDIS results will open a new chapter probing general
principles of QCD TMD-formalism. In this review main focus is given to the very recent results obtained by the COMPASS
collaboration from first ever multi-dimensional extraction of SIDIS transverse spin asymmetries.
\end{Abstract}
\vfill
\begin{Presented}
CIPANP 2015: Twelfth Conference on the Intersections of Particle and Nuclear Physics\\
Vail (CO), USA, May 19-24, 2015 \\
\end{Presented}
\vfill
\end{titlepage}
\def\thefootnote{\fnsymbol{footnote}}
\setcounter{footnote}{0}
%


%
%
%
\section{Introduction}	
In past decades, measurements and following study of the spin dependent and \emph{unpolarized} azimuthal effects in
SIDIS has became a priority direction in experimental and theoretical high-energy physics. Using standard SIDIS
notations, the differential cross-section can be written in single-photon exchange approximation as
\cite{Kotzinian:1994dv}--\cite{Diehl:2005pc}:
{\small
\begin{eqnarray}\nonumber
&& \hspace*{-1.4cm}\frac{{d\sigma }}{{dxdydzp_{T}^{h}dp_{T}^{h}d{\phiH}d\phiS }} = 2\left[ {\frac{\alpha }{{xy{Q^2}}}\frac{{{y^2}}}{{2\left( {1 - \varepsilon } \right)}}\left( {1 + \frac{{{\gamma ^2}}}{{2x}}} \right)} \right]\left( {{F_{UU,T}} + \varepsilon {F_{UU,L}}} \right) \\ \nonumber
%
%
&&\hspace*{-1.4cm} \times\Bigg\{ 1 + \sqrt {2\varepsilon \left( {1 + \varepsilon } \right)} \textcolor[rgb]{0.00,0.07,1.00}{A_{UU}^{\cos {\phi _h}}}\cos {\phiH} + \varepsilon \textcolor[rgb]{1.00,0.00,0.00}{A_{UU}^{\cos 2{\phi _h}}}\cos \left( {2{\phiH}} \right) + \lambda \sqrt {2\varepsilon \left( {1 - \varepsilon } \right)} \textcolor[rgb]{0.00,0.07,1.00}{A_{LU}^{\sin {\phi _h}}}\sin {\phiH}\\ \nonumber
&&\hspace*{-1.1cm}+\,{{S}_{T}}\Big[\textcolor[rgb]{1.00,0.00,0.00}{A_{UT}^{\sin \left( {{\phiH} - {\phiS}} \right)}}\sin \left( {{\phiH} - {\phiS}} \right) + \varepsilon \textcolor[rgb]{1.00,0.00,0.00}{A_{UT}^{\sin \left( {{\phiH} + {\phiS}} \right)}}\sin \left( {{\phiH} + {\phiS}} \right) + \varepsilon \textcolor[rgb]{1.00,0.00,0.00}{A_{UT}^{\sin \left( {3{\phiH} - {\phiS}} \right)}}\sin \left( {3{\phiH} - {\phiS}} \right)\\ \nonumber
&&\hspace*{+0.1cm}+\,\sqrt {2\varepsilon \left( {1 + \varepsilon } \right)} \textcolor[rgb]{0.00,0.07,1.00}{A_{UT}^{\sin {\phiS}}}\sin {\phiS} + \sqrt {2\varepsilon \left( {1 + \varepsilon } \right)} \textcolor[rgb]{0.00,0.07,1.00}{A_{UT}^{\sin \left( {2{\phiH} - {\phiS}} \right)}}\sin \left( {2{\phiH} - {\phiS}} \right)\Big]\\ \nonumber
&&\hspace*{-1.1cm}+\,{{S}_{T}}\lambda \Big[\sqrt {\left( {1 - {\varepsilon ^2}} \right)} \textcolor[rgb]{1.00,0.00,0.00}{A_{LT}^{\cos \left( {{\phiH} - {\phiS}} \right)}}\cos \left( {{\phiH} - {\phiS}} \right)\\
&&\hspace*{+0.1cm}+\,\sqrt {2\varepsilon \left( {1 - \varepsilon } \right)} \textcolor[rgb]{0.00,0.07,1.00}{A_{LT}^{\cos {\phiS}}}\cos {\phiS} + \sqrt {2\varepsilon \left( {1 - \varepsilon } \right)} \textcolor[rgb]{0.00,0.07,1.00}{A_{LT}^{\cos \left( {2{\phiH} - {\phiS}} \right)}}\cos \left( {2{\phiH} - {\phiS}} \right)\Big]\Bigg\},
\label{eq:SIDIS}
\end{eqnarray}
}
with $\varepsilon = (1-y -\frac{1}{4}\slim \gamma^2 y^2)/(1-y +\frac{1}{2}\slim y^2 +\frac{1}{4}\slim \gamma^2
y^2)$\footnote{ratio of longitudinal and transverse photon fluxes}, $\gamma = 2 M x/Q$.
 Target transverse polarization (${S}_{T}$)
\footnote{in reality polarization vector has a small longitudinal component w.r.t. virtual photon direction which leads
to small deviations of amplitudes (not discussed in this letter, for details see \cite{Parsamyan:2013ug})}
dependent part of this general expression contains eight
azimuthal modulations in the $\phi_h$ and $\phi_S$ azimuthal angles of the produced hadron and of the nucleon spin,
correspondingly. Each modulation leads to a $A_{BT}^{w_i(\phiH, \phiS)}$ Transverse-Spin-dependent Asymmetry (TSA)
defined as a ratio of the associated structure function $F_{BT}^{w_i(\phiH,\phiS)}$ to the azimuth-independent one
${F_{UU}}={{F_{UU,T}} + \varepsilon {F_{UU,L}}}$. Here the superscript of the asymmetry indicates corresponding
modulation, the first and the second subscripts - respective ("U"-unpolarized,"L"-longitudinal and "T"-transverse)
polarization of beam and target. Five amplitudes which depend only on ${S}_{T}$ are the target Single-Spin Asymmetries
(SSA), the other three which depend both on ${S}_{T}$ and $\lambda$ (beam longitudinal polarization) are known as
Double-Spin Asymmetries (DSA).
%

In the QCD parton model approach four out of eight transverse spin asymmetries have Leading Order (LO) or leading
"twist" interpretation and are described by the convolutions of twist-two Transverse-Momentum-Dependent (TMD) Parton
Distribution Functions (PDFs) and Fragmentation Functions (FFs) \cite{Kotzinian:1994dv}--\cite{Mulders:1995dh}.

The first two: $A_{UT}^{sin(\phi_h-\phi_S)}$ "Sivers" and $A_{UT}^{sin(\phi_h+\phi_S)}$ "Collins" effects
 \cite{Adolph:2012sn,Adolph:2012sp} are the most studied ones. These asymmetries are given as convolutions of:
 $f_{1T}^{\perp q}$ Sivers PDF convoluted with $D_{1q}^h$ ordinary FF and $h_{1}^{q}$ "transversity" PDF convoluted
 with the $H_{1q}^{\perp h}$ Collins FF, respectively. The other two LO terms are the $A_{UT}^{\sin(3\phiH -\phiS )}$
 single-spin asymmetry (related to $h_{1T}^{\perp\,q}$ ("pretzelosity") PDF
 \cite{Parsamyan:2014uda}--\cite{Parsamyan:2007ju}) and $A_{LT}^{\cos (\phiH -\phiS )}$ DSA (related to $g_{1T}^q$
("worm-gear") distribution function
\cite{Parsamyan:2014uda}--\cite{Parsamyan:2007ju,Kotzinian:2006dw,Anselmino:2006yc}).
{\small
\begin{align}\label{eq:LO_as}
&A_{UT}^{\sin (\phiH -\phiS )} \propto f_{1T}^{\bot q} \otimes
D_{1q}^h,\ \
A_{UT}^{\sin (\phiH +\phiS )} \propto h_1^q \otimes H_{1q}^{\bot
h},  \\
&A_{UT}^{\sin (3\phiH
-\phiS )} \propto h_{1T}^{\bot q} \otimes H_{1q}^{\bot
h},\ A_{LT}^{\cos (\phiH -\phiS )} \propto g_{1T}^q \otimes
D_{1q}^h.\nonumber
\end{align}
}

Remaining four asymmetries 
are so-called "higher-twist" effects\footnote{in equation \ref{eq:SIDIS} the twist-2 amplitudes are marked in red and
higher-twist ones in blue}. Corresponding structure functions enter at sub-leading order ($Q^{-1}$) and contain terms
given as various mixtures of twist-two and twist-three (induced by quark-gluon correlations) parton distribution and
fragmentation functions \cite{Bacchetta:2006tn,Mao:2014aoa,Mao:2014fma}. However, applying wildly adopted so-called
"Wandzura-Wilczek approximation" this higher twist objects can be simplified to twist-two level (see
\cite{Bacchetta:2006tn,Mulders:1995dh} for more details):
{\small
\begin{align}\label{eq:NLO_as}
&A_{UT}^{\sin (\phiS )} \propto {Q}^{-1}({h_1^q \otimes
H_{1q}^{\bot h} +f_{1T}^{\bot q} \otimes D_{1q}^h }),\ \nonumber \\
&A_{UT}^{\sin (2\phiH -\phiS )} \propto
{Q}^{-1}({h_{1T}^{\bot q} \otimes H_{1q}^{\bot h}
+f_{1T}^{\bot q} \otimes D_{1q}^h }),\  \\
&A_{LT}^{\cos (\phiS )} \propto {Q}^{-1}(g_{1T}^q \otimes
D_{1q}^h),\ \
A_{LT}^{\cos (2\phiH -\phiS )} \propto {Q}^{-1}
(g_{1T}^q \otimes D_{1q}^h).\nonumber
\end{align}
}

In general, TSAs being convolutions of different TMD functions are known to be complex objects \textit{a priori}
dependent on the choice of kinematical ranges and multidimensional kinematical phase-space. Thus, ideally, asymmetries
have to be extracted as multi-differential functions of kinematical variables in order to reveal the most complete
multivariate dependence. In practice, available experimental data often is too limited for such an ambitious approach
and studying dependence of the asymmetries on some specific kinematic variable one is forced to integrate over all the
others sticking to one-dimensional approach.
Presently, one of the hottest topics in the field of spin-physics is the study of TMD evolution of various PDFs and FFs
and related asymmetries. Different models predict from small up to quite large $\sim1/Q^2$ suppression of the QCD-evolution
effects attempting to describe available experimental observations and make predictions for the future
ones \cite{Aybat:2011ta,Echevarria:2014xaa,Sun:2013hua}. Additional experimental measurements exploring different
$Q^2$ domains for fixed $x$-range are necessary to further constrain the theoretical models.
The work described in this review is a unique and first ever attempt to explore behaviour of TSAs in the multivariate
kinematical environment.
For this purpose COMPASS experimental data was split into five different $Q^2$ ranges giving an opportunity to study
asymmetries as a function of $Q^2$ at fixed bins of $x$. Additional variation of $z$ and $p_T$ cuts allows to deeper
explore multi-dimensional behaviour of the TSAs and their TMD constituents.
\section{Multidimensional analysis of TSAs}	
The analysis was carried out on COMPASS data collected in 2010 with transversely polarized proton data. General event
selection procedure as well as asymmetry extraction and systematic uncertainty definition techniques applied for this
analysis are identical to those used for recent COMPASS results on Collins, Sivers and other TSAs
\cite{Adolph:2012sn}--\cite{Parsamyan:2007ju}.

The eight target transverse spin dependent "raw" asymmetries are extracted simultaneously from the fit using extended
unbinned maximum likelihood method and then are corrected for average depolarization factors ($\varepsilon$-depending
factors in equation \ref{eq:SIDIS} standing in front of the amplitudes), dilution factor and target and beam (only
DSAs) polarizations evaluated in the given kinematical bin \cite{Adolph:2012sn}--\cite{Parsamyan:2007ju}.
{\small
\begin{eqnarray}\label{eq:depol}
    && D^{\sin(\phiH -\phiS )}(y) \cong 1,\;\;
    D^{\cos(\phiH -\phiS )}(y) = \sqrt {\left( {1 - \varepsilon^{2} } \right)} \approx \frac {y(2-y)} {1+(1-y)^2},\nonumber\\
    && D^{\sin(\phiH +\phiS )}(y) = D^{\sin(3\phiH -\phiS )}(y)
 = \varepsilon \approx \frac {2(1-y)} {1+(1-y)^2}, \\
    && D^{\sin(2\phiH -\phiS )}(y) = D^{\sin(\phiS )}(y) = \sqrt {2\varepsilon \left( {1 + \varepsilon } \right)} \approx \frac
{2(2-y)\sqrt{1-y}} {1+(1-y)^2}, \nonumber \\
    &&D^{\cos(2\phiH -\phiS )}(y) = D^{\cos(\phiS )}(y) = \sqrt {2\varepsilon \left( {1 - \varepsilon } \right)} \approx \frac
{2y\sqrt{1-y}}
    {1+(1-y)^2}.\nonumber
\end{eqnarray}
}

Primary sample is defined by the following standard DIS cuts: $Q^2>1$ $(GeV/c)^2$, $0.003<x<0.7$ and $0.1 <y < 0.9$ and
two more \textit{hadronic} selections: $p_T>0.1$ GeV/c and $z>0.1$.

\begin{figure}[h]
\includegraphics[width=7.8cm]{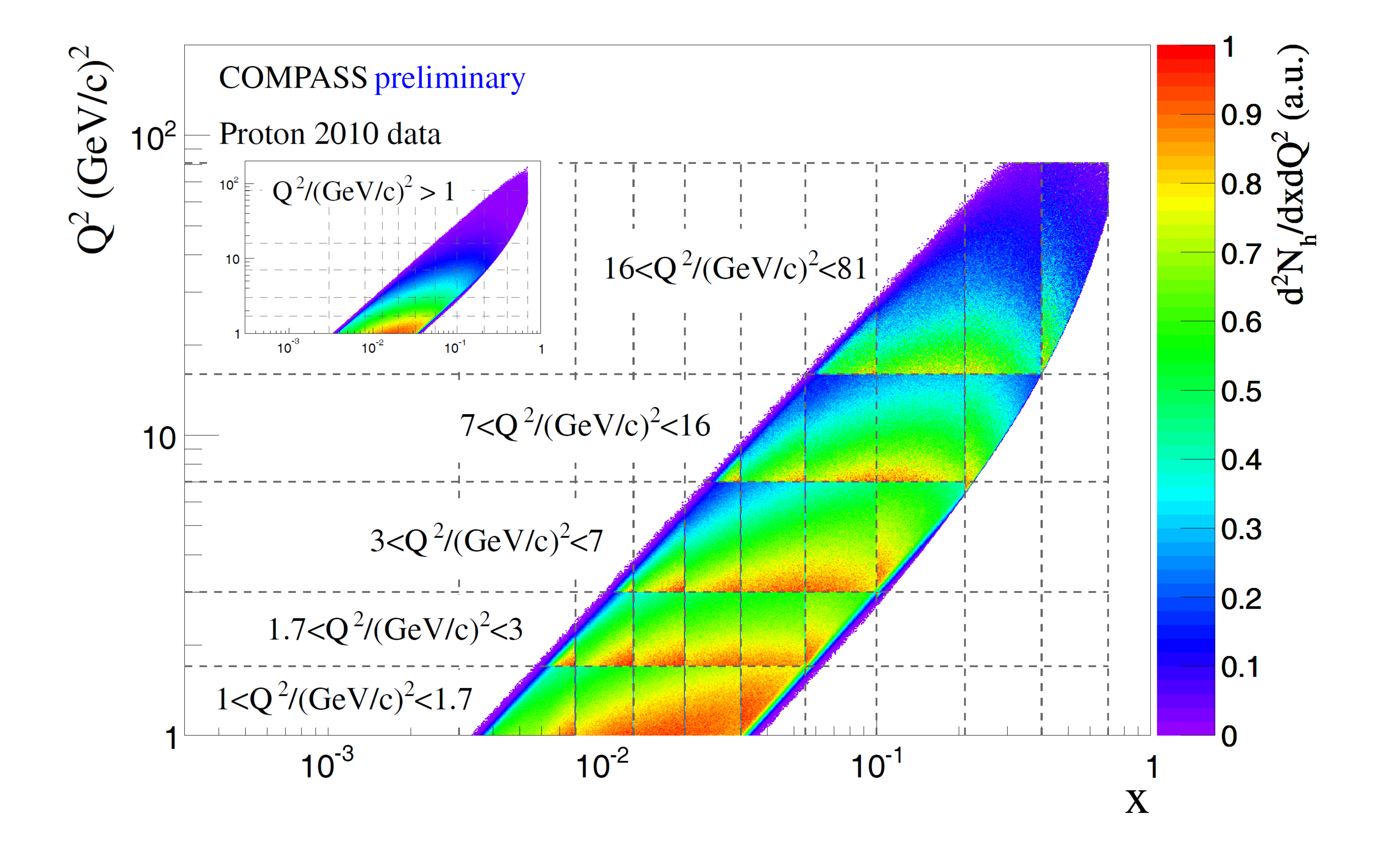}
\includegraphics[width=7.8cm]{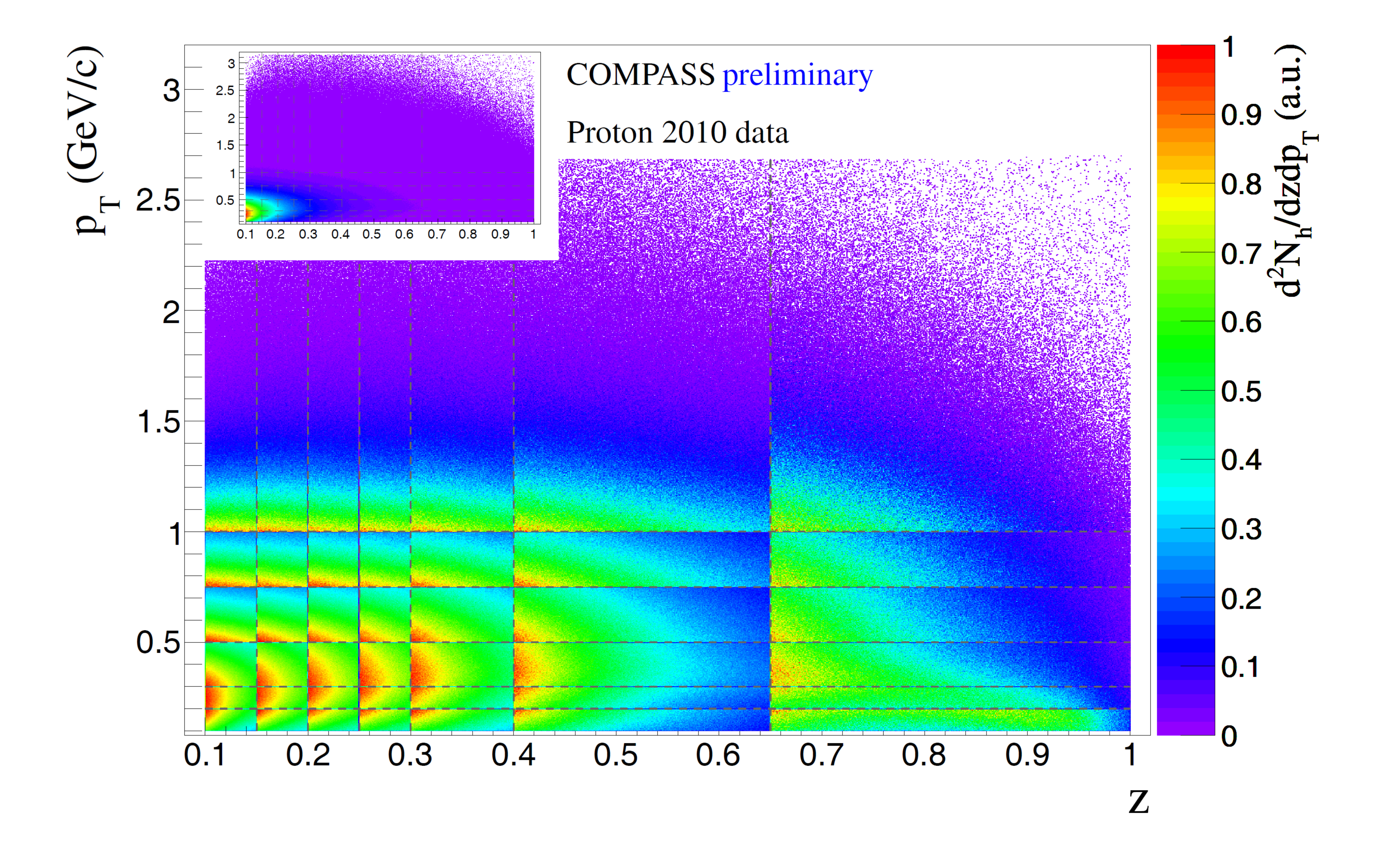}
\caption{COMPASS $x:Q^2$ (left) and $z$:$p_T$ (right) phase space coverage. \label{f1}}
\end{figure}

In order to study possible $Q^2$-dependence the $x$:$Q^2$ phase-space covered by COMPASS experimental data has been
divided into $5\times9$ two-dimensional grid (see left plot in Figure~\ref{f1}). Selected five $Q^2$-ranges are the
following ones: $Q^{2}/(GeV/c)^2$ $\in$ $[1;1.7],[1.7;3],[3;7],[7;16],[16;81]$ \footnote{$16<Q^{2}/(GeV/c)^2<81$
selection repeats the definition of the so-called "high-mass" range:most promising domain for future COMPASS--Drell-Yan
TSA-analyses \cite{Parsamyan:2014uda,Parsamyan:2015cfa}.}.
In addition, each of this samples has been divided into five $z$ and five $p_T$ (GeV/c) sub-ranges defined as follows:\\
$z>0.1$, $z>0.2$, $0.1<z<0.2$, $0.2<z<0.4$ and $0.4<z<1.0$\\
$p_T>0.1$, $0.1<p_T<0.75$, $0.1<p_T<0.3$, $0.3<p_T<0.75$ and $p_T>0.75$.
Using various combinations of aforementioned cuts and ranges, asymmetries have been
extracted for following "3D" and "4D" configurations: 1) $x$-dependence in $Q^2$-$z$ and $Q^2$-$p_T$ grids.
2) $Q^2$-dependence in $x$-$z$ and $x$-$p_T$ grids. 3) $Q^2$- (or $x$-) dependence in $x$-$p_T$ (or $Q^2$-$p_T$)
grids for different choices of $z$-cuts.
Another approach was used to focus on $z$- and $p_T$-dependences in different $x$-ranges.
 For this study the two-dimensional $z$:$p_T$ phase-space has been divided into $7\times6$ grid as it is demonstrated
in right plot in Figure~\ref{f1}. Selecting in addition three $x$-bins: $0.003<x<0.7$, $0.003<x<0.032$, $0.032<x<0.7$
asymmetries have been extracted in "3D: $x$-$z$-$p_T$" grid. In the next section several examples of COMPASS
preliminary results obtained for multi-dimensional target transverse spin dependent azimuthal asymmetries are
presented.

\section{Results}
\begin{figure}[H]
\includegraphics[width=15.4cm]{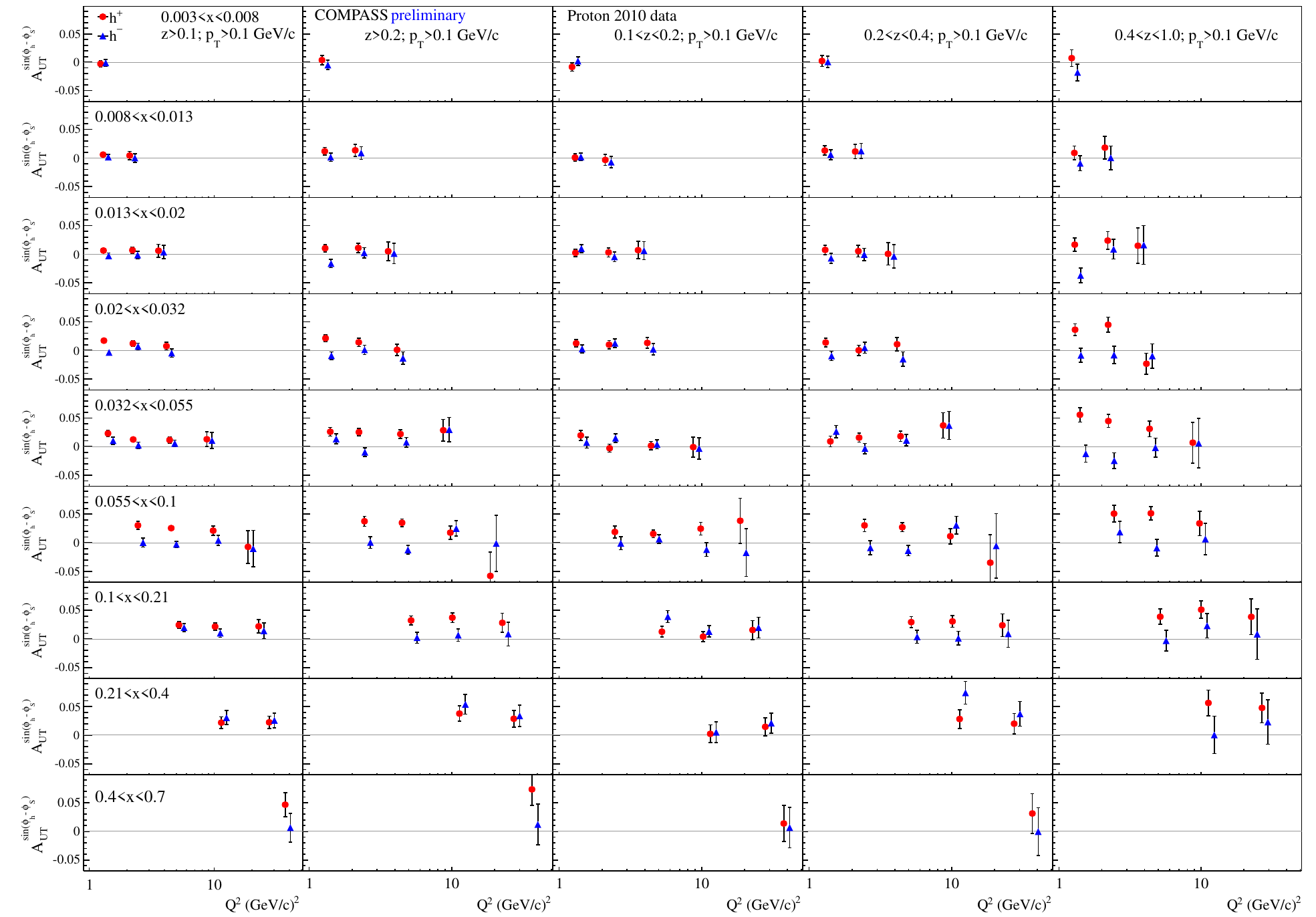}
\caption{Sivers asymmetry in "3D": $Q^2$-$p_T$-$x$ (top) and $x$-$z$-$Q^2$ (bottom). \label{f2}}
\end{figure}

\footnote{Discussed results have been first presented at the SPIN-2014 conference \cite{Parsamyan:2015dfa},
see also \cite{Parsamyan:QCDEV15,Parsamyan:DSPIN15}.}
As an example of "3D" Sivers effect, results for the extracted $x$-$z$-$Q^2$ configuration are presented in the
Figure~\ref{f2}. The results shown at the plot serve as a direct input for TMD-evolution related studies. In fact, in
several x-bins there are some hints for possible decreasing $Q^2$-dependence for positive hadrons which become more
evident at large $z$. As a general observation, sizable Sivers asymmetry tending to increase with $z$ and $p_T$ was
observed for positive hadrons, while for negative hadrons there are some indications for a positive signal at
relatively large $x$ and $Q^2$ and negative effect at low $x$.

\begin{figure}[H]
\includegraphics[width=15.4cm]{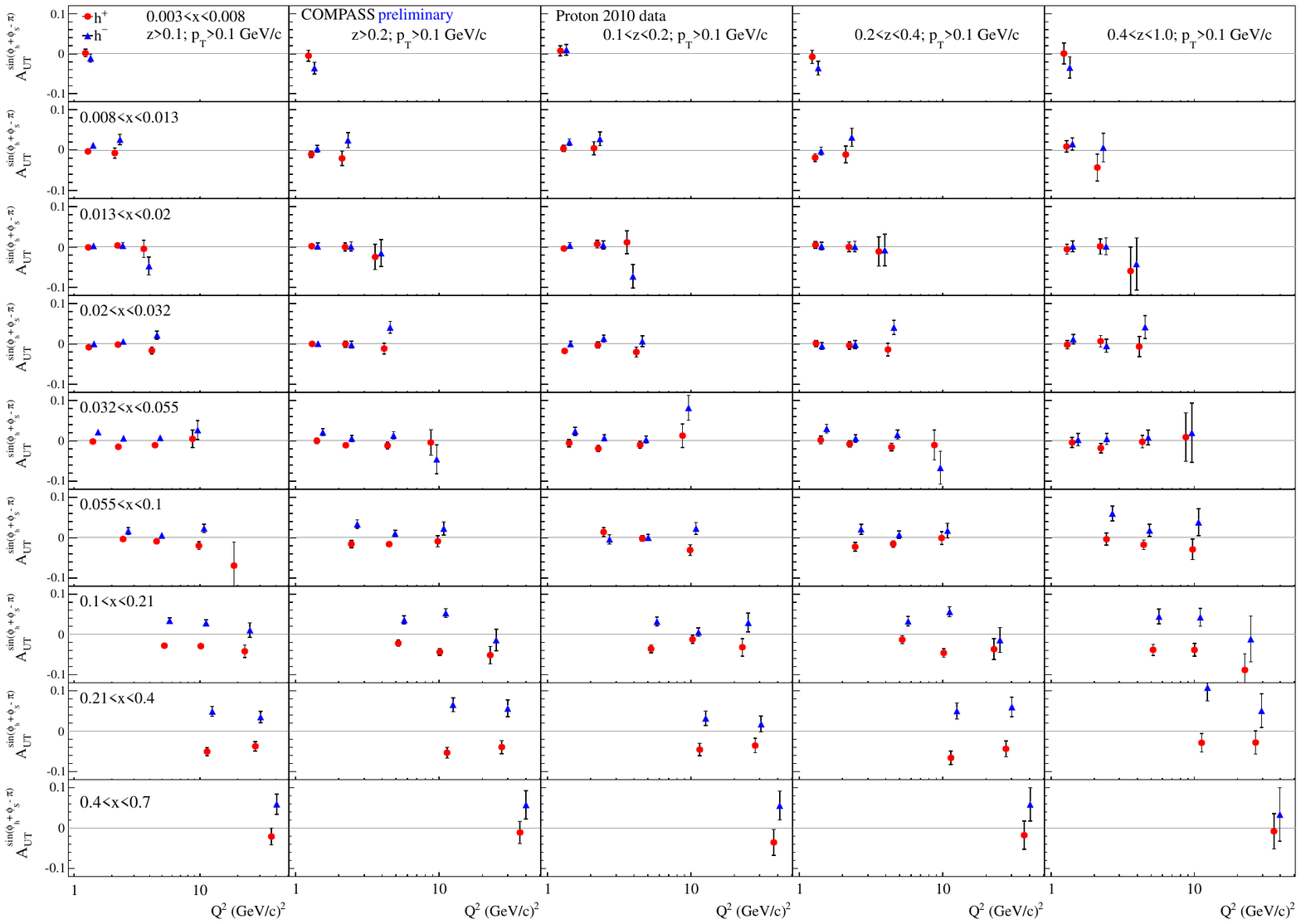}
\includegraphics[width=15.6cm]{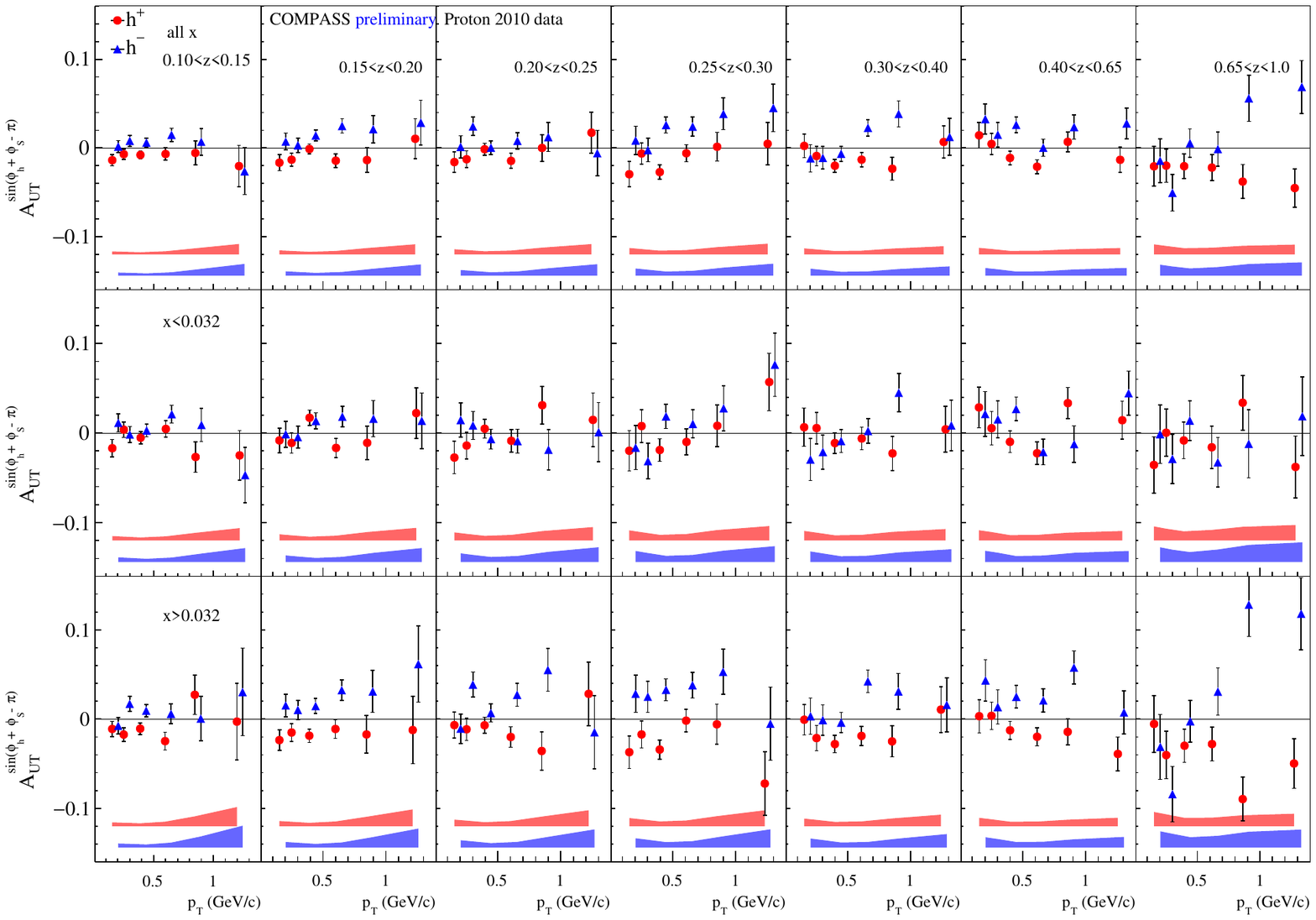}
\vspace*{-0.7cm}
\caption{Collins asymmetry in "3D": $x$-$z$-$Q^2$" (top) and "$x$-$z$-$p_T$" (bottom). \label{f3}}
\end{figure}

\begin{figure}[H]
\hspace*{-10pt}
\includegraphics[width=15.6cm]{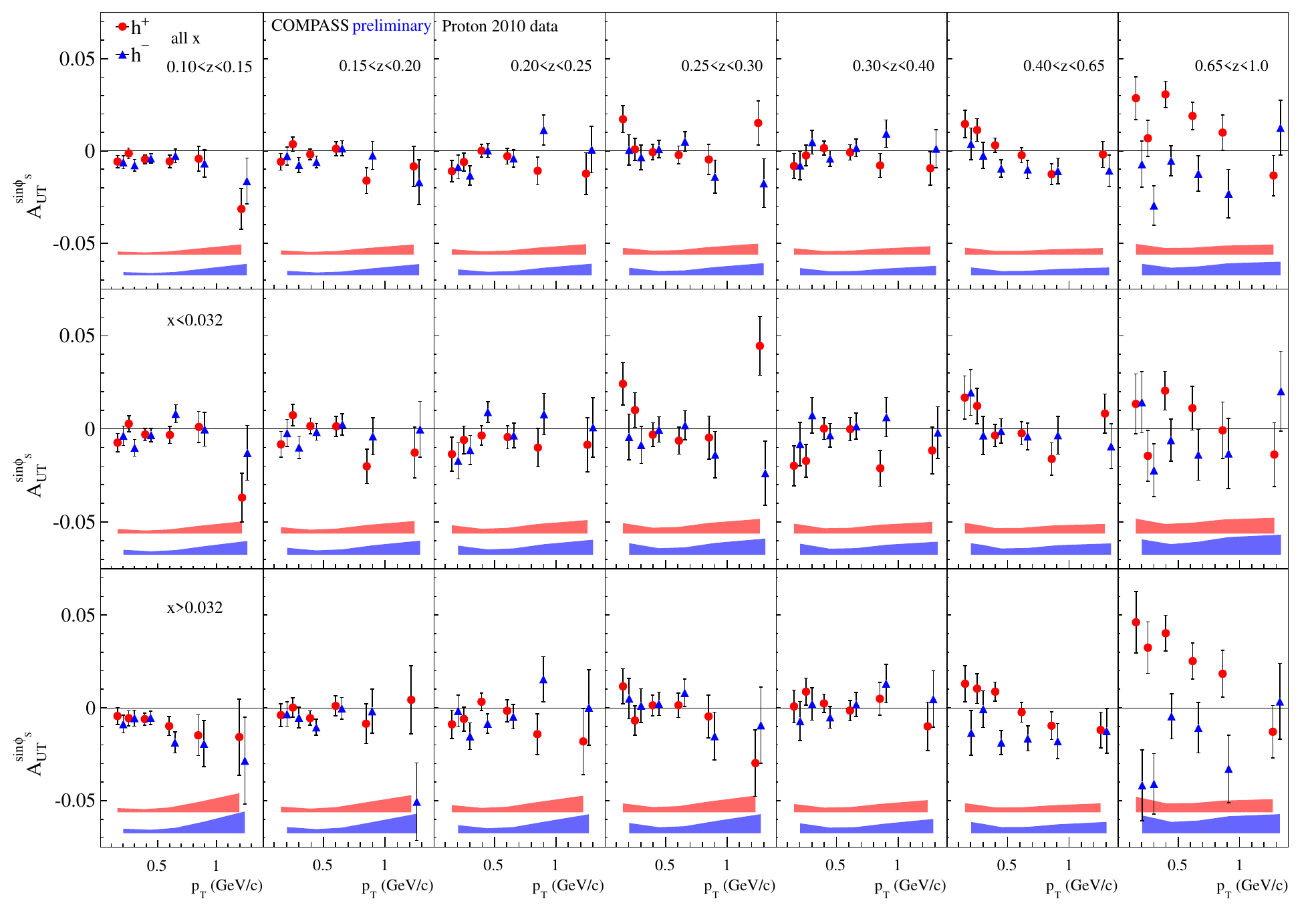}
\includegraphics[width=15.2cm]{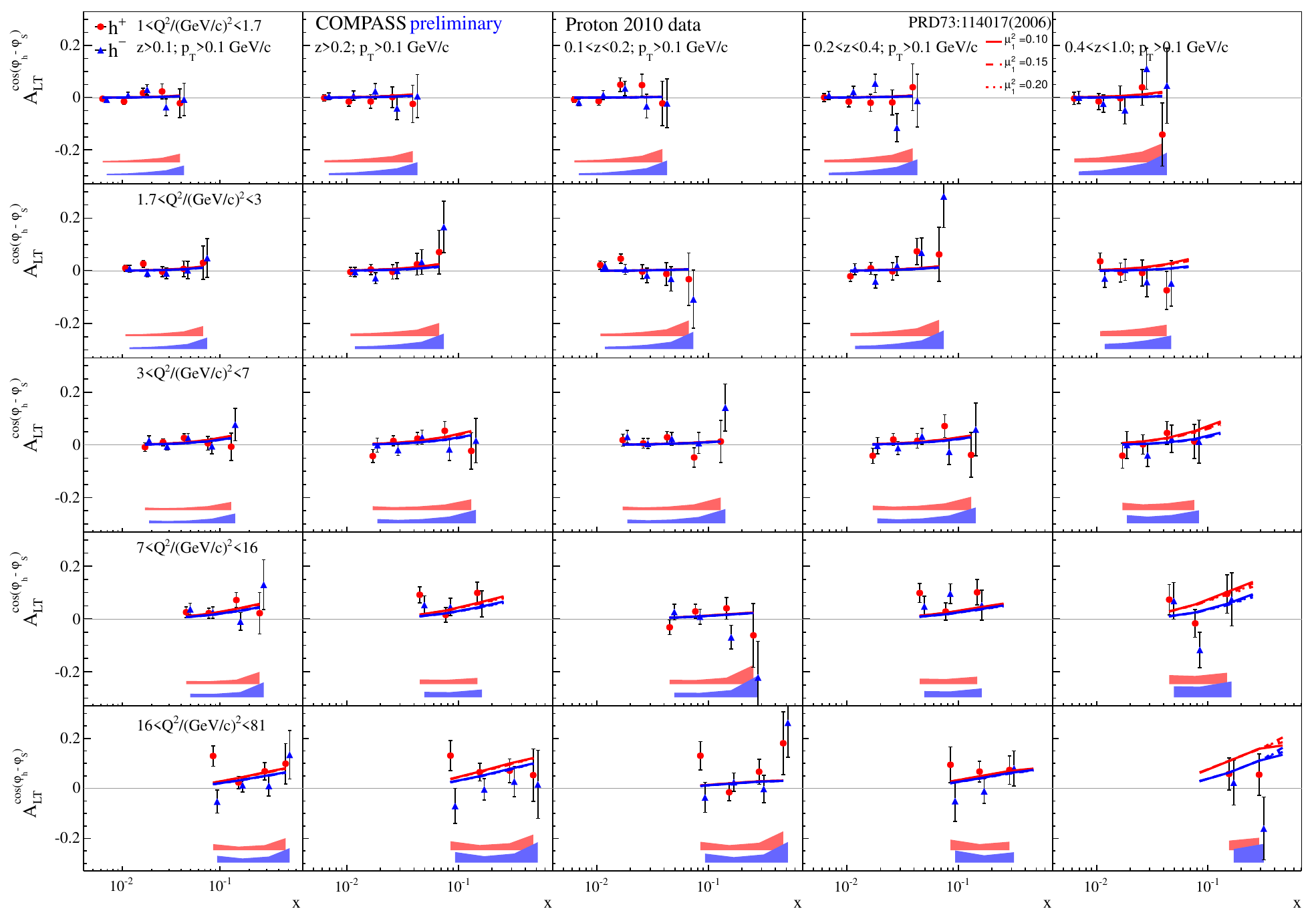}
\vspace*{-0.3cm} \caption{Top: $A_{UT}^{\sin (\phi _s )}$ asymmetry in "3D" ($x$-$z$-$p_T$). Bottom: $A_{LT}^{\cos
(\phiH -\phiS )}$ in "3D" ($Q^2$-$z$-$x$) superimposed with theoretical predictions from [13].
\label{f4}}
\end{figure}

In Figure~\ref{f3} Collins asymmetry is shown in "3D: $x$-$z$-$Q^2$". Clear "mirrored" behaviour for positive and
negative hadron amplitudes is being observed in most of the bins. Amplitudes tend to increase in absolute value with
both $z$ and $p_T$. There are no clear indications for $Q^2$-dependence.
Another SSA which is found to be non-zero at COMPASS is the $A_{UT}^{\sin (\phi _s )}$ term which is presented in
Figure~\ref{f4} (top) in "3D: $x$-$z$-$p_T$" configuration. Here the most interesting is the large $z$-range were
amplitude is measured to be sizable and non zero both for positive and negative hadrons.

The bottom plot in the Figure~\ref{f4} is dedicated to the $A_{LT}^{\cos (\phiH -\phiS )}$ DSA explored in "3D:
$Q^2$-$z$-$x$" grid and superimposed with the theoretical curves from \cite{Kotzinian:2006dw}. This is the only DSA
which appears to be non-zero at COMPASS and the last TSA for which a statistically significant signal has been
detected. Remaining four asymmetries are found to be small or compatible with zero within available statistical
accuracy which is in agreement with available predictions \cite{Mao:2014aoa,Mao:2014fma,Lefky:2014eia}.
\section{Conclusions}
The first ever multidimensional extraction of the whole set of target transverse spin dependent azimuthal asymmetries
has been done at COMPASS with proton data collected in 2010. Various multi-differential configurations has been tested
exploring $x$:$Q^2$:$z$:$p_T$ phase-space. Particular attention was given to probes of possible $Q^2$-dependence of
TSAs, serving a direct input to TMD-evolution related studies. Several interesting observations have been made studying
the results obtained for Sivers,
 Collins, $A_{LT}^{cos(\phi_h-\phi_S)}$ and $A_{UT}^{sin(\phi_S)}$ asymmetries. Other four asymmetries were found to be
compatible with zero within given statistical accuracy. This highly differential data set for the eight asymmetries,
combined with past and future relevant data obtained by other collaborations will give a unique opportunity to access
the whole set of TMD PDFs and test their multi-differential nature and key features.
\vspace*{-2.0cm}
\vspace*{-2.0cm}
%
%

%
%

%

\end{document}